 \documentclass[%
 reprint,superscriptaddress,
 amsmath,amssymb,prx,longbibliography]{revtex4-1}
\usepackage[final]{changes}

\usepackage{physics}
\usepackage{xcolor}
\usepackage{bm}
\usepackage{xtab,afterpage,longtable}
\usepackage{lipsum}
\usepackage{graphicx}
\usepackage{dcolumn}
\usepackage{booktabs} 
\usepackage{multirow}
\usepackage{color}
\usepackage{eucal}
\usepackage{mathrsfs}  
\usepackage{siunitx}
\usepackage{makecell}
\usepackage{hyperref}

\setaddedmarkup{\textcolor{red}{#1}}

\begin{document}

\title{Cooperative quantum optical effects of moir\'{e} exciton superlattices}

\author{Haowei Xu}%
\email{haoweixu@cityu.edu.hk}
\affiliation{Department of Physics, City University of Hong Kong, Kowloon, Hong Kong SAR, China }

\author{Wang Yao}%
\affiliation{New Cornerstone Science Laboratory, Department of Physics, The University of Hong Kong, Hong Kong, China}


\author{Ju Li}%
 \affiliation{Department of Nuclear Science and Engineering, Massachusetts Institute of Technology, Cambridge, MA 02139, USA}
 \affiliation{Department of Materials Science and Engineering, Massachusetts Institute of Technology, Cambridge, MA 02139, USA}


\begin{abstract}
The unique properties of two-dimensional moir\'{e} systems have been widely studied from many perspectives. However, relatively little work has investigated how the \emph{real space} structure of moir\'{e} systems can directly engender novel properties and functionalities. In this work, we explore how intriguing cooperative optical effects can emerge from the real-space superlattice formed by moir\'{e} excitons. Particularly,  we show that the collective moir\'{e} exciton states can have either strongly enhanced (superradiant) or suppressed (subradiant) radiative decay rate, depending on their in-plane wavevector. These super- and subradiant states can be efficiently switched by a gate-induced electric field gradient. Moreover, the cooperative transmittance $T$ of the atomically thin moir\'{e} system can be switched from $T\approx 0$ (opaque) to $T\approx 1$ (transparent) with  $\lesssim 1~\%$ heterostrain or a $1^{\circ}$ adjustment in the twist angle $\theta$. These features are robust against non-radiative losses and \deleted{inhomogeneity} \added{disorders}, making the moir\'{e} system a highly versatile platform for cooperative quantum optics with potential applications in e.g., single photon storage and switching.

\end{abstract}

\maketitle

\textbf{Introduction.} When two-dimensional materials are stacked with small twist angle and/or lattice mismatch, the moir\'{e} pattern is formed, which exhibits a wide variety of exotic phenomena, such as correlated insulating states, unconventional
superconductivity~\cite{cao2018unconventional,cao2018correlated}, and orbital ferromagnetism~\cite{sharpe2019emergent,chen2020tunable,serlin2020intrinsic}. The photonic and optoelectronic properties of moir\'{e} systems~\cite{baek2020highly,du2023moire}, especially those involving moir\'{e} excitons~\cite{tran2019evidence,seyler2019signatures,jin2019observation}, have also garnered considerable interest~\cite{baek2020highly,du2023moire} with implications in tunable quantum emitters~\cite{yu2017moire,baek2020highly}, infrared detectors~\cite{ma2022intelligent}, topological excitons~\cite{wu2017topological}, etc.

Most previous studies on optical responses have focused on the properties derived from a \emph{single} localized moir\'{e} exciton. In reality, however, mori\'{e} excitons form a super-lattice. One might argue that the total optical response of the superlattice is a simple summation of the responses from each  moir\'{e} exciton. 
This viewpoint aligns with conventional treatments of optical responses in condensed-matter systems, where one usually defines a susceptibility $\xi$ per unit cell. The total response is then taken as $N\xi$ with $N$ the number of unit cells participating in the light-matter interaction.  This approach is generally valid provided that the phase-matching condition of different light beams is properly included~\cite{shen1983principles}. Within such a framework, the real-space structure of the system enters only through the overall scaling factors $N$, consistent with the common view that reciprocal $k$-space structure plays a more central role than real-space structure in condensed-matter physics. 

Nevertheless, such a treatment may \emph{not} be adequate for moir\'{e} excitons. While the real-space structure of moir\'{e} systems is relatively simple (e.g., honeycomb-like), few studies have examined how this structure can have \emph{direct} consequences for their properties~\cite{wang2025emergent,zheng2025forster}. Crucially, for optical responses, the moir\'{e} superlattice constant $a_M$ \added{($10\sim 100~\rm nm$ when the twist angle is $\theta\lesssim 1^{\circ}$)} can  \deleted{be comparable to} \added{become a considerable fraction of} the resonant wavelength $\lambda_0$ (hundreds of nanometers)\deleted{when the twist angle is $\theta\lesssim 1^{\circ}$}. In this regime, moir\'{e} excitons can interfere constructively or destructively during light-matter interactions, so their real-space structure may fundamentally reshape the overall optical response, which can be far from the simple summation $N\xi$. 

The discussions above motivate us to investigate the \emph{cooperative} optical response of the moir\'{e} exciton superlattices. Cooperative optical effects, such as collective Lamb shift~\cite{glicenstein2020collective} and the subradiant optical mirror effect~\cite{rui2020subradiant}, have primarily been explored in e.g., artificial atom arrays~\cite{bettles2016enhanced,asenjo2017exponential,shahmoon2017cooperative}, but have rarely been studied in intrinsic condensed-matter systems. 
While isolated atoms benefit from narrow non-radiative linewidth, it is experimentally challenging to assemble large and perfectly ordered atomic arrays, limiting their practical applicability. In contrast, moir\'{e} systems naturally form ordered lattices with unit filling, that is, each lattice site hosts one exciton. This is an arrangement that is difficult to realize in artificial atomic platforms. Moreover, the moir\'{e} superlattice constant $a_M$ and many other properties can be conveniently controlled by the twist angle $\theta$ and heterostrain. 
These features make moir\'{e} systems highly promising platforms for exploring cooperative quantum optical phenomena and their potential applications.

In the following, we begin by introducing the properties of moir\'{e} excitons and show that each exciton can be modeled as a two-level system with an optical transition dipole. We then explore several significant cooperative quantum optical effects of moir\'{e} exciton superlattices. Particularly, we show that collective moir\'{e} exciton states can have extremely high (superradiant) or low (subradiant) radiative decay rates, depending on their in-plane wavevector. The subradiant states can have lifetimes extended by orders of magnitude compared with a single exciton, making them promising candidates for photon memory~\cite{zhang2024realization,liu2025millisecond}.  Additionally, a gate-induced electric field gradient can dynamically switch the system between superradiant and subradiant states, enabling efficient photon storage and retrieval. We further demonstrate that though the moir\'{e} system is atomically thin, its light transmittance $T$ can be tuned from $T\approx 1$ to $T\approx 0$,  with less than $2~\%$ heterostrain or $1^{\circ}$ change in the twist angle, allowing the system to function as an efficient single photon switch~\cite{sun2018single}. These properties remain robust against non-radiative losses and inhomogeneous broadening up to a few hundred GHz, well within experimental reach, given that the overall linewidth of moir\'{e} excitons can be as narrow as 20 GHz~\cite{seyler2019signatures}.
\begin{figure}
    \centering
    \includegraphics[width=0.9\linewidth]{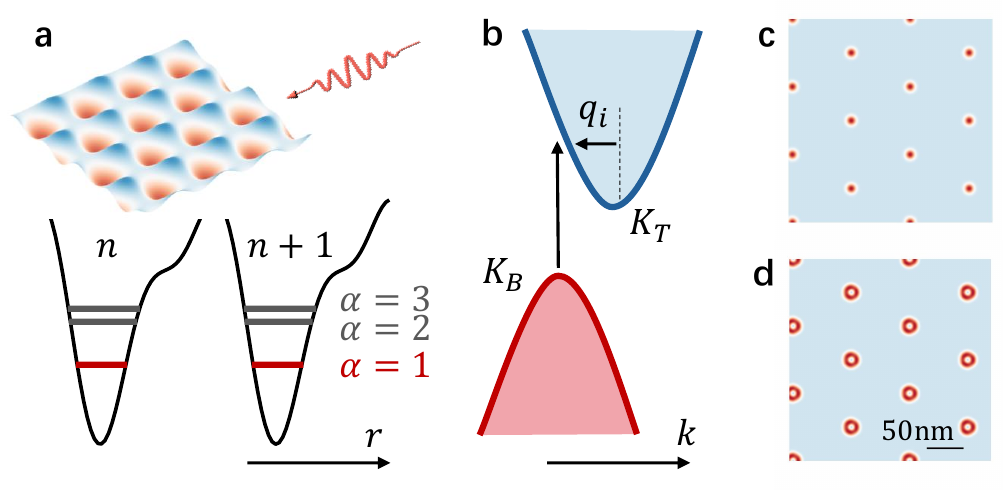}
    \caption{(a) Illustrations of moir\'{e} potential and exciton lattice in real space. Upper panels show a two-dimensional moir\'{e}  potential landscape. Lower panel shows multiple localized moir\'{e}  exciton states in each potential minimum, labeled by $\alpha$. We focus on the lowest energy $\alpha=1$ state. (b) Illustration of the electronic band structure of the moir\'{e} system in the unfolded Brillouin zone. The electron acquires a wave vector $q_i$ relevant to $K_T$ after vertical optical transition from $K_B$. There are three symmetry-equivalent $q_{i=1,2,3}$, related by $2\pi/3$ rotations. (c, d) Real space wavefunction $\vert \langle r\vert \chi_{\alpha} \rangle \vert$ of (a) $\alpha=1$ and (b) $\alpha=2$ exciton with $\theta=0.2^{\circ}$. Blue (red) color denotes small (large) wavefunction amplitude. }
    \label{fig:illustration}
\end{figure}

\textbf{Properties of moir\'{e} excitons}. We first introduce the properties of single moir\'{e} excitons, focusing on their spatial localization and optical transitions. We will show that moir\'{e} exciton on each superlattice site can be modeled as a two-level system with ground state $\vert G \rangle$, excited state $\vert R \rangle$, and transition dipole moment $d$ (Figure~\ref{fig:points_interactions}). Detailed discussions can be found in \added{Sections 1-4} of the Supplementary Materials~\cite{supplementary}, which includes Refs.~\cite{trolle2014theory,yu2017moire,wu2018theory,ahn2020low,jiang2018microsecond,robert2020measurement,taghizadeh2018gauge,taghizadeh2020nonlinear,louca2023interspecies,seyler2019signatures,asenjo2017exponential,garcia2007colloquium}.

Without the moir\'{e} potential, a free exciton state can be labeled by $\ket{Q}$, with total momentum $Q$ and total energy $\hbar\omega_0 + \frac{\hbar Q^2}{2M}$. Here, $\hbar\omega_0$ is the energy of a zero-momentum exciton, typically on the order of $1~\rm eV$~\cite{wang2018colloquium}. Meanwhile, $\frac{\hbar Q^2}{2M}$ is the kinetic energy of the exciton with $M$ the total mass of the electron and hole. Notably, $\vert Q \rangle$ are delocalized plane-wave-like states, and the exciton center-of-mass coordinate $r$ has a simple spatial profile $Q(r) \sim e^{iQ \cdot r}$.  The optical transition dipole between the ground state $\vert G\rangle$ (i.e., no exciton) and $\vert Q\rangle$ is nonzero only when $Q = q_{i=1,2,3}$, where $q_1 = K_T - K_B$, while $q_2$ and $q_3$ are obtained from $q_1$ by $2\pi/3$ and $4\pi/3$ rotations, respectively. Here $K_T$ and $K_B$ are the $K$-valley of the top and bottom layers (Figure~\ref{fig:illustration}b and Figure S2 in Ref.~\cite{supplementary}). This is because the photon momentum is negligible, so an electron at $K_B$ would acquire a wavevector of $q_i$ relative to $K_T$ after a vertical optical transition (Figure~\ref{fig:illustration}b). The optically bright momenta $Q = q_1, q_2, q_3$ are equivalent due to the three-fold rotational symmetry of the system. 

The moir\'{e} structure leads to a potential field  $\Delta (r)$ for the excitons, which has the same periodicity as the moir\'{e} supercell (Figure~\ref{fig:illustration}a). This is because at different locations of the moir\'{e} supercell, the upper and lower layers have different stacking patterns, resulting in different local environments. The potential $\Delta (r)$ results in the hybridization between free exciton states, namely $\bra{Q} \Delta(r) \ket{Q'} = \sum_{n=1}^6 V_n \delta_{Q, Q'+b_n}$, where $\delta$ is the Kronecker delta, $b_n$ are the moir\'{e} reciprocal lattice vectors, and $V_n$ characterizes the depth of the moir\'{e} potential, which is typically tens of meV for inter-layer excitons~\cite{yu2017moire,wu2017topological}.  

The new moir\'{e} exciton eigenstates are $\ket{\chi_{\alpha, Q}} = \sum_{Q'} f^{\alpha}_{QQ'} \ket{Q'}$, where $f^{\alpha}_{QQ'}$ is the wavefunction. Here, $Q$ is restricted to the first moir\'{e} Brillouin zone, 
while other $Q$-points outside the first moir\'{e} Brillouin zone are folded back into it, giving rise to an additional band index $\alpha$.
The energy difference between different $\alpha$ states is on the order of meV (Figure~\ref{fig:illustration}a, Section 2 in Ref.~\cite{supplementary}). This is analogous to the formation of electron bands in the nearly free electron model~\cite{kittel2018introduction}.  We focus on the lowest-energy $\alpha=1$ mode that is optically bright. \added{The influence of other modes is discussed in Section 4.2 in Ref.~\cite{supplementary}.} Due to the hybridization induced by $\Delta(r)$, the optically bright moir\'{e} exciton eigenstates contain equivalent contributions from $Q=q_1, q_2, q_3$ modes, resulting in a total momentum of zero (since $q_1 + q_2 + q_3 = 0$). In the following, we will use the shorthand $\vert \chi \rangle$ to denote this particular mode unless otherwise noted. 



Next, we consider the spatial profile of the moir\'{e} exciton.  The real-space wavefunction, expressed as $\langle r \vert \chi \rangle = \sum_{Q'} f_{QQ'} e^{iQ'\cdot r}$, is localized at the minima of the moir\'{e} potential (Figure~\ref{fig:illustration}c,~\ref{fig:illustration}d). One can define approximate exciton wave packets $\ket{R} = \sum_{Q'} W (Q') \ket{Q'}$, mimic an exciton localized at the lattice site $R$ with a proper choice of $W(Q')$. 
Notably, the width $w$ of the moir\'{e} exciton wavefunction is on the order of a few nanometers, much smaller than the optical wavelength, which is hundreds of nanometers. Therefore, each localized moir\'{e} exciton can be treated as a point-like dipole in optical response calculations. This significantly simplifies the analyses of their cooperative optical response.

The width of the wave packet $w$ is also crucial in the optical transitions. Particularly, the optical transition dipole moment $d \equiv \bra{G}\hat{d}\ket{R}$ is proportional to $w$. Here $\hat{d}$ is the dipole operator and $\vert G\rangle$ is the ground state, i.e., no excitons at this lattice point. The property of  $d\propto w$ is similar to the optical transition dipoles of quantum dots, which scale linearly with their size~\cite{jacak2013quantum}. For a wide range of twist angles, one has $w\sim 0.3 a_M \propto 1/\theta$ . Hence, we adopt $d  \approx \frac{d_0}{\theta} $ with $d_0 = 0.1~e\cdot \si\angstrom$. As $d$ increases for larger $a_M$ (smaller $\theta$), the dipole-dipole interaction among moir\'{e} excitons can remain strong even if $a_M$ is large (Section 2 in Ref.~\cite{supplementary}). 

Strictly speaking, excitons are bosons, so multiple excitons can, in principle, occupy the same potential minimum.  However, if more than one exciton occupies the same site, then the on-site exciton-exciton interactions would introduce a resonance frequency shift~\cite{yu2017moire} (analogous to the Hubbard $U$  for electrons), much larger than the energy scale relevant here. Hence, when the incident light is (nearly) resonant with the transition between $\vert G \rangle$ (no exciton) and $\vert R\rangle$ (one exciton), states with multiple excitons are far off-resonance and can be neglected. \added{In Section 4 in Ref.~\cite{supplementary}, we will systematically analyze the validity of using the two-level system approximation for the moir\'e exciton super-lattice.}


\begin{figure}
    \centering
    \includegraphics[width=0.35\linewidth]{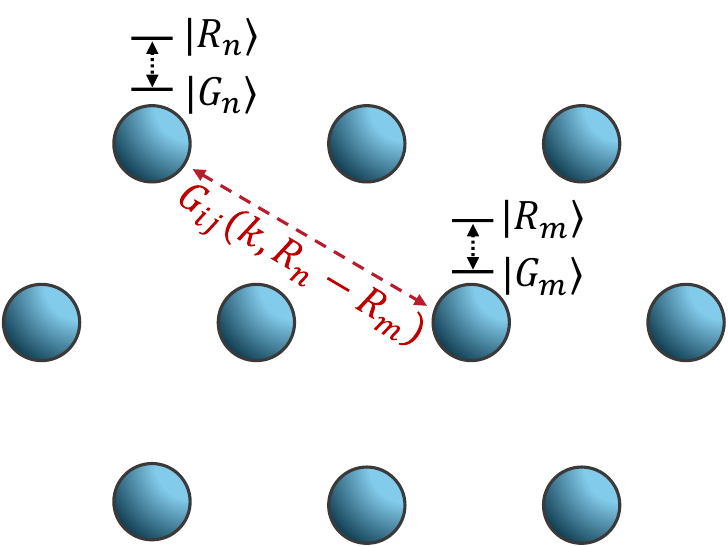}
    \caption{An ordered  moir\'{e}  superlattice. Each lattice site $R_n$ is modeled by a two-level system $\vert R_n \rangle$ and $\vert G_n \rangle$ with transition dipole $d$. The red arrow indicates the {electromagnetic} interactions among  moir\'{e} excitons, characterized by the Green's function $G_{ij}$.}
    \label{fig:points_interactions}
\end{figure}

\textbf{Interactions among moir\'{e} excitons}. In the previous section, we focused on the properties of a single moir\'{e} exciton. Now we turn to the ordered superlattice formed by $N$ moir\'{e} excitons. We will use a subscript $n$ to denote an exciton at lattice site $R_n$. Each lattice site $R_n$ is thus modeled by a two-level system $\vert R_n \rangle$ and $\vert G_n \rangle$ with transition dipole $d$ (Figure~\ref{fig:points_interactions}).

These localized moir\'{e} excitons interact with one another through dipole-dipole interactions. For two excitons at locations $R_n$ and $R_m$, one has the Green's function as 
\begin{equation}\label{eq:Greens_function}
\begin{aligned}
    G_{ij}(k,r) = \frac{e^{ikr}}{4\pi k^2 r^3} [ & (k^2r^2 + ikr - 1)\delta_{ij} \\
    & + (-k^2r^2 - 3ikr +3) \frac{r_i r_j}{r^2} ]
    \end{aligned}
\end{equation}
Here $k$ is the wavevector of the electromagnetic field. We defined $r = R_n - R_m$. $i,j$ are Cartesian indices,  and $\delta_{ij}$ is the Kronecker delta. The vacuum Green's function~\cite{novotny2012principles} is used here for simplicity. \added{The influence of the dielectric environment arising from, for example, the substrate and encapsulation layers, is discussed in Section 8 in Ref.~\cite{supplementary}.}


The interactions among dipoles, encoded in the Green's function, can lead to exotic cooperative effects, as we will discuss below.

\begin{figure}
    \centering
    \includegraphics[width=1\linewidth]{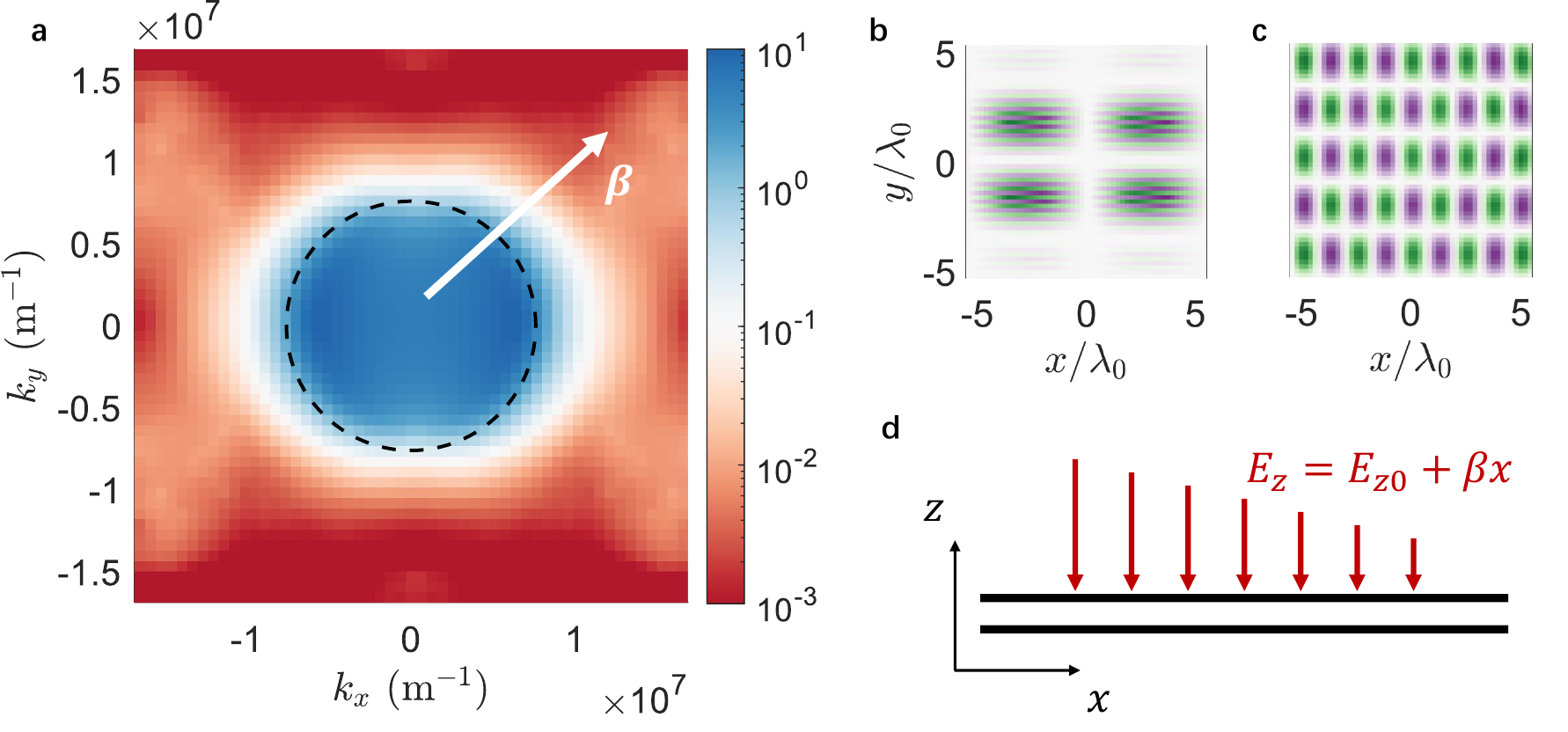}
    \caption{(a) Collective radiative decay rate $\Gamma_{\nu}$ as a function of in-plane wavevector $k_{\parallel}$ for a finite lattice with approximately $50\times 50$ excitons. The dashed circle denotes the light cone with $\vert k_{\parallel} \vert = k_0$.    
    (b, c) Real part of the wavefunction of the collective exciton in real space, i.e., $\mathrm{Re}[\langle \psi_n \vert \nu(k_{\parallel})\rangle]$. (b) and (c) correspond to two states with $\vert k_{\parallel}\vert > k_0$ and $\vert k_{\parallel}\vert < k_0$. Each pixel corresponds to moir\'{e} lattice site $R_n$. Blue (purple) color indicate positive (negative) values of the wavefunction.  (d) A collective exciton state can be switched between super- to sub-radiant states by an electric field gradient $\beta$. The electric field is along out-of-plane $z$ direction, while the gradient is along the in-plane $x$ or $y$ direction.}
    \label{fig:super-sub-radiance}
\end{figure}

\textbf{Superradiant and subradiant states}. We will first discuss super- and sub-radiance of the collective moir\'{e} exciton states~\cite{dicke1954coherence,asenjo2017exponential,masson2020many}, and how they can be dynamically switched, providing a mechanism for light storage and retrieval.

The total Hamiltonian of the exciton lattice is~\cite{asenjo2017exponential}
\begin{equation}\label{eq:H_eff_dipole_dipole}
\begin{aligned} 
& H_{\rm eff}  = \hbar \omega_0  \sum_{n=1}^N \vert R_n \rangle \langle R_n \vert \\
& + \frac{k_0^2 d^2}{\varepsilon_0} \sum_{m,n}  G_{xx}(k_0, R_n-R_m) 
  \vert G_n \rangle  \vert R_m \rangle  \langle R_n \vert  \langle G_m \vert 
\end{aligned}
\end{equation}
The first term is the non-interacting Hamiltonian, which is zero ($\hbar \omega$) if a lattice site is on $\ket{G_n}$ (on $\vert R_n\rangle$). The second term describes the long-range dipole-dipole interaction among moir\'{e} excitons. Here, $\varepsilon_0$ is vacuum permittivity, and we define $k_0 = \omega_0/c_0$ with $c_0$ the speed of light. The transition dipole $d$ is assumed to be the same for all excitons. This term accounts for Forster-type non-radiative coupling that allows moir\'{e} excitons to effectively hop between the trapping sites~\cite{zheng2025forster}. We did not include the out-of-plane permanent dipole $d_z \equiv \langle R_n \vert \hat{d} \vert R_n\rangle$ of the inter-layer exciton, as it is much smaller than $d$~\cite{yu2017moire}. Without loss of generality, we consider excitons excited by light polarized along the $x$-direction, so only $G_{xx}$ appears in Eq.~(\ref{eq:H_eff_dipole_dipole}). The full $H_{\rm eff}$ has a dimension of  $2^N$, but we focus on the single-excitation subspace spanned by the basis $\ket{\psi^n} \equiv \ket{G_1 G_2,\cdots, R_n, \cdots, G_N}, n = 1,2, \cdots, N$, whose dimension is $N$. \added{The validity of using only the single-excitation subspace will be discussed in Section 4 in Ref.~\cite{supplementary}. In essence, in terms of the interaction described in Eq.~(\ref{eq:H_eff_dipole_dipole}), including the double- and higher-excitation states would not affect the properties within the single-excitation subspace.}


The effective Hamiltonian $H_{\rm eff}$ is non-Hermitian. Nevertheless, one can still investigate its eigenstates $\ket{\nu}$ with complex eigenvalues $J_{\nu} - i \frac{\Gamma_{\nu}}{2}$. Here, $J_{\nu}$ and $\Gamma_{\nu}$ correspond to the collective lineshift and radiative decay rate, respectively. Generally, $\ket{\nu}$ are collective exciton states, similar to the Bloch functions, and $\vert \langle \psi_n\vert \nu \rangle\vert^2$ is the probability of finding the exciton at lattice site $R_n$. 

For an infinite exciton lattice, $\ket{\nu}$ can be labeled by the in-plane wavevector $k_{\parallel}$, according to Bloch's theorem. That is, one has $\vert \nu(k_{\parallel}) \rangle \propto \sum_n e^{i k_{\parallel}\cdot R_n} \vert \psi_n \rangle$. Notably, for $k_{\parallel} > k_0$, one has $\Gamma_{\nu} = 0$, and the excitons are completely dark (Section \added{5} in Ref.~\cite{supplementary}). For $k_{\parallel} < k_0$, superradiance $\Gamma_{\nu} > \gamma_0$ occurs. This superradiance-subradiance transition arises from momentum conservation: the in-plane momentum $k_{\parallel}$ of the emitted photon must match that of the collective exciton. For $k_{\parallel}  > k_0$,  the out-of-plane wavevector $\sqrt{k_0^2 - k_{\parallel}^2}$ becomes imaginary, producing an evanescent field and suppressing radiation ($\Gamma_{\nu}=0$).

In practice, the moir\'{e} exciton lattice is finite, and one cannot assign a definite $k_{\parallel}$ to each eigenstate. In this regard, we numerically diagonalize $H_{\rm eff}$ for a \added{$N=2874\approx $} $50\times 50$ exciton lattice, and selected wavefunctions $\ket{\nu}$ are plotted in Figures~\ref{fig:super-sub-radiance}b,~\ref{fig:super-sub-radiance}c. These states exhibit semi-periodic patterns, allowing for a Fourier transform to extract the dominant ${k}_{\parallel}$ component, which can be used to label $\vert \nu \rangle$ as $\langle \psi_n \vert \nu(k_{\parallel}) \rangle \approx e^{ik_{\parallel}\cdot R_n}$. Using this approach, we obtained the $\Gamma_{\nu}$-$k_{\parallel}$ relationship (Figure~\ref{fig:super-sub-radiance}a) for a finite lattice. Notably, there is still a transition from superradiance ($\Gamma_{\nu} \gg \gamma_0$) to subradiance ($\Gamma_{\nu} \ll \gamma_0$)  at the light cone boundary $\vert k_{\parallel} \vert = k_0$. The transition is smooth due to the finite-size effect, and states with $\vert k_{\parallel} \vert  < k_0$ are not entirely dark - their $\Gamma_{\nu}$ is small but non-zero. \deleted{Note that the sharpness of the transition depends on the number of lattice sites, as in other phase transitions.} \added{Here, the $50 \times 50$ lattice has a spatial size of about $10~\mathrm{\mu m} \times 10~\mathrm{\mu m}$. This system size is chosen for two main reasons: 1) it is already significantly larger than the wavelength of the resonant light ($\lesssim 1~\rm \mu m$), ensuring that collective effects are well developed; 2) experimentally, fabricating substantially larger high-quality moir\'e samples may be challenging. It should be noted that the contrast between superradiant and subradiant states is stronger for larger lattice size, and becomes strongest for an infinite lattice, as shown in Section 5.3 in Ref.~\cite{supplementary}.}


\textbf{Switching between superradiant and subradiant states}. A collective exciton state excited by an external photon must have $k_{\parallel} < k_0$ and is therefore superradiant. This allows the photon to be efficiently transferred into an exciton state. For long-term photon storage, however, the excitation should reside in a subradiant state, which has a much longer lifetime.

Hence, effective light storage and retrieval require controllable switching between superradiant and subradiant states. This can be achieved by applying a gate electric field with a spatial gradient, which leads to a Hamiltonian $H_{\beta} = \sum_n  d_z \cdot \beta \cdot R_n \vert R_n \rangle \langle R_n \vert$, where $d_z = \langle R_n \vert \hat{d} \vert R_n\rangle$ is the out-of-plane dipole of the exciton. This electric field is along the $z$ direction, while the gradient $\beta$ lies in-plane (along $x$ or $y$ in Figure~\ref{fig:super-sub-radiance}d). $H_{\beta}$ induces a position-dependent phase, effectively shifting the exciton wavevector, namely~\cite{plankensteiner2015selective} 
\begin{equation}\label{eq:electric_field_gradient} 
\begin{aligned}
k_{\parallel} \to k_{\parallel} + \frac{d_z \cdot \beta}{\hbar} \tau,
\end{aligned}
\end{equation}
with the $\tau$ the duration of the applied electric field gradient. With a suitable $\tau$, a superradiant exciton can be pushed beyond the light-cone and become subradiant. In practice, this can be achieved within nanoseconds with $\beta\sim 10^{10}~\rm V/m^2$. To restore the superradiant state, one can simply reverse the direction of the electric field ($-\beta$) for the same duration $\tau$. Note that $\beta$ does not need to be uniform in real space. 

Intuitively, Eq.~(\ref{eq:electric_field_gradient}) resembles the semiclassical equation of motion for an electron in a crystal, $k\to k + eE\tau/\hbar$, where the electron charge $e$ couples directly to the electric field $E$. In contrast, an exciton carries no net charge but only a dipole moment $d_z$, so its wavevector can be modified only via the electric-field gradient $\beta$.

\begin{figure}
    \centering
    \includegraphics[width=1.0\linewidth]{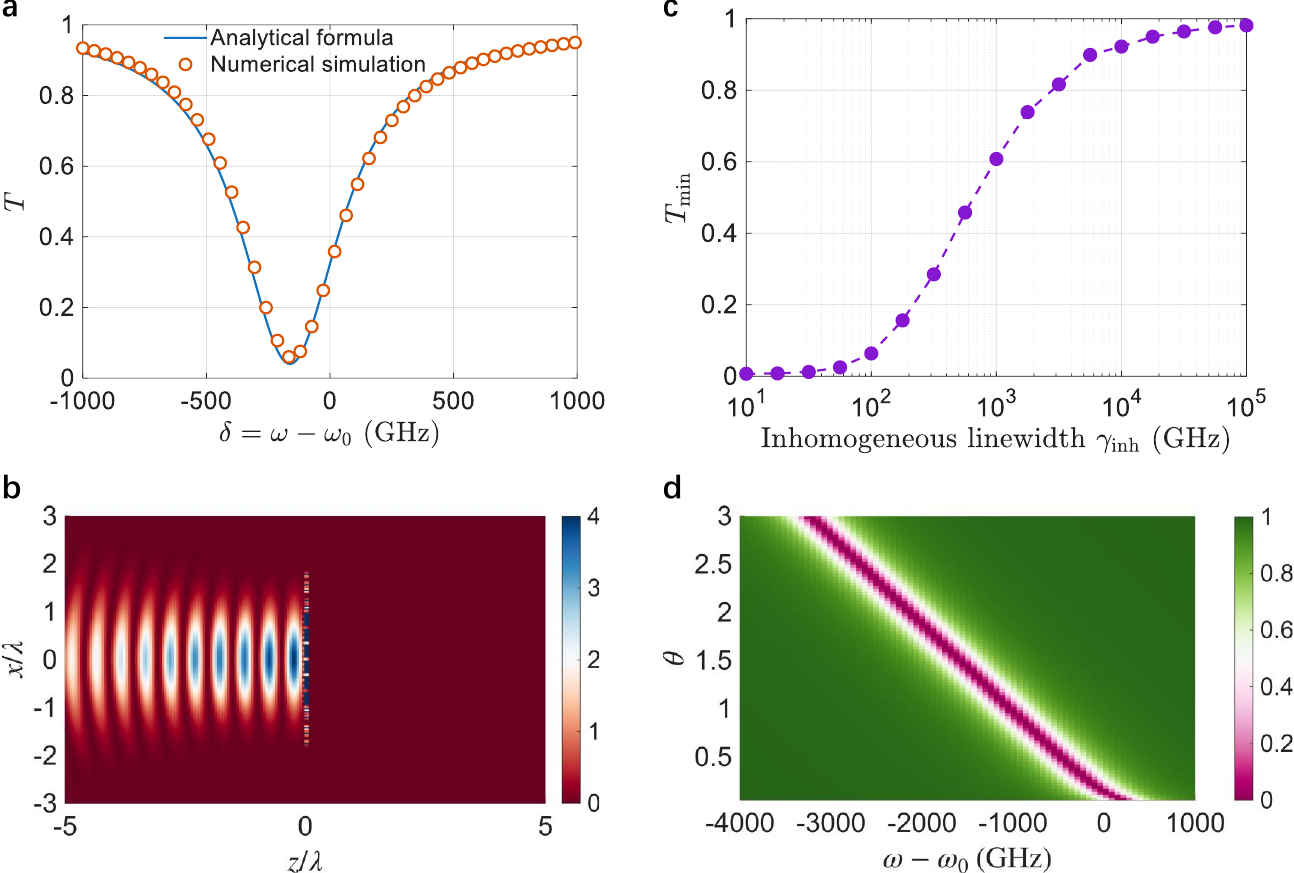}
    \caption{(a) Transmittance $T(\omega)$ as a function of frequency detuning $\omega-\omega_0$ with $\theta = 0.2^{\circ}$. (b) The electric field profile $\vert E(r) \vert^2$ at $\omega-\omega_0\approx 161~\rm GHz$, where $T(\omega)$ reaches the minimum at $\theta = 0.2^{\circ}$.  Light is incident from left, and the fields $\vert E(r) \vert^2$ are normalized by $E_0^2$ with $E_0$ the incident field strength.  (c) Minimum of $T$ as a function of inhomogeneous linewidth with $\theta=0.2^{\circ}$.  (d) $T$ as a function of $\omega-\omega_0$ for varied $\theta$. The non-radiative linewidth is taken as $\gamma_{\rm nr} = 100~\rm GHz$.}
    \label{fig:transmittance}
\end{figure}

\textbf{Cooperative photon scattering.} Besides photon absorption and emission, cooperative optical effects also appear in photon scattering. As an example, we study the transmittance $T$ of the moir\'{e} exciton lattice. Typically, few-layer 2D materials have $T\approx 1$, as they are too thin to significantly reflect or absorb light. For example,  monolayer graphene has $T\approx 0.97$~\cite{zhu2014optical}. Remarkably, as we show below, the moir\'{e} exciton lattice can exhibit nearly zero transmittance. 

Assuming an incident field $E_0(r)$ with frequency $\omega$ and wavelength $\lambda$, the total field $E(r)$ after scattering with the excitons should satisfy~\cite{bettles2016enhanced,shahmoon2017cooperative,garcia2007colloquium}
\begin{equation}\label{eq:total_field_formal_solution}
\begin{aligned}
E_i(r) = E_{0,i}(r) + \frac{4\pi^2 \alpha}{\varepsilon_0\lambda^2} \sum_{jn} G_{ij} (k, r-R_n) E_j(R_n)
\end{aligned}
\end{equation}
For simplicity, we consider $x$-polarized light incident normally along $z$ onto the exciton lattice plane ($x$-$y$ plane). Other geometries can be analyzed similarly. Eq.~(\ref{eq:total_field_formal_solution}) indicates that the total field $E(r)$ is the sum of the incident field $E_0(r)$ and the field scattered by the exciton dipole at $R_n$, mediated by the propagator $G_{ij} (k, r-R_n) $. The exciton polarizability is $\alpha(\omega) = \frac{2d^2 \omega_0}{\hbar} \frac{1}{\omega_0^2 - \omega^2 - i \gamma \omega}$. The total linewidth of a single exciton is given by $\gamma = \gamma_r+ \gamma_{\rm nr}$, where $\gamma_r$ and $\gamma_{\rm nr}$ are the radiative and non-radiative components, respectively. Each moir\'{e} exciton may also experience an inhomogeneous shift in its resonance frequency $\omega_0$, characterized by an inhomogeneous linewidth $\gamma_{\rm inh}$.


We first numerically study a finite exciton lattice with $\theta = 0.2^{\circ}$ ($a_M \approx 0.11\lambda_0$) with $\gamma_{\rm nr}=100~\rm GHz$  and $\gamma_{\rm inh}= 0$  (details in \added{Section 6} in Ref.~\cite{supplementary}). The transmittance $T$ versus frequency detuning $\delta = \omega - \omega_0$ is plotted in Figure~\ref{fig:transmittance}a. Intriguingly, nearly zero transmission appears at $\delta \approx -161~\rm GHz$. The field profile (Figure~\ref{fig:transmittance}b) shows that the scattered field and the incident field form a standing wave in the $z<0$ region. In the $z>0$ region, the two fields add destructively, leading to zero transmission (\added{Section 6.1} in Ref.~\cite{supplementary}). 

Analytically, for an infinite exciton lattice with $a_M < \lambda$, one has (\added{Section 6.2} in Ref.~\cite{supplementary})
\begin{equation}\label{eq:r_analytical} 
\begin{aligned}
T(\omega) = \left\vert 1 - \frac{ \frac{i}{2}(\frac{\lambda_0^3}{\lambda^3}  \gamma_{0}  + \Gamma )}{\omega - \omega_0 + \Delta + \frac{i}{2}(\gamma_{0} + \gamma_{\rm nr} + \Gamma)} \right\vert^2,
\end{aligned}
\end{equation}
where the collective lineshift $\Delta$ and the collective radiative decay rate $\Gamma$ are
\begin{equation}\label{eq:cooperative_reflectance} 
\begin{aligned}
\Delta  = \frac{3 \gamma_{0}}{2} \frac{\lambda_0^3}{\lambda^3} \lambda \mathrm{Re}\{\mathcal{G}_{xx}(0)\}, \,\,
\Gamma  = \frac{3 \gamma_{0}}{4\pi} \frac{\lambda_0^3}{\lambda^3}  \frac{\lambda^2}{A}  - \frac{\lambda_0^3}{\lambda^3}  \gamma_{0},\\
\end{aligned}
\end{equation}
with $\mathcal{G}_{xx}(0) = \sum_{n\neq 0} G_{xx}(k, R_n)$ and $A$ the area of the moir\'{e} superlattice. Intuitively, the moir\'{e} excitons interact cooperatively [Eq.~(\ref{eq:Greens_function}) and Figure~\ref{fig:points_interactions}], forming a collective excitation that behaves like a giant atom.  Zero transmittance occurs at the collective resonance $\Delta+\omega - \omega_0=0$ when $\gamma_{\rm nr} = 0$, which is demonstrated in Figure~\ref{fig:transmittance}d for varied $\theta$. 

Notably, this cooperative effect is robust to inhomogeneity $\gamma_{\rm inh}$ and non-radiative damping $\gamma_{\rm nr}$, provided that they remain below the \emph{collective} radiative decay rate $\Gamma$ in Eq.~(\ref{eq:r_analytical}).  Our results indicate that transmittance below $0.2$ can be realized with $\gamma_{\rm inh}, \gamma_{\rm nr} \lesssim 100~\rm GHz$ (Figure~\ref{fig:transmittance}c and \deleted{Section 5.3} \added{Section 6.3} in Ref.~\cite{supplementary}). 

More importantly, this effect is highly sensitive to the real space structure \deleted{twist angle $\theta$} \added{the real-space structure of the moir\'e exciton superlattice}, enabling an efficient light switch. At a fixed light frequency, nearly zero transmission  $T\approx 0$ and full transmission  $T\approx 1$ can be toggled by varying \added{the twist angle} $\theta$ by less than $1^{\circ}$ (Figure~\ref{fig:transmittance}d). This is because the collective shift $\Delta$ strongly depends on the moir\'{e} lattice spacing, which is controlled by the twist angle. From this perspective, a heterostrain $u$ can also be used to tune $T$. This is because applying heterostrain is effectively equivalent to tuning $\theta$ in terms of modifying the lattice constant $a_M \propto \frac{1}{\sqrt{u^2 + \theta^2}}$. Using heterostrain $u\lesssim 1\%$, one can realize in situ and reversible control over $T$. 

\textbf{Discussions.} It is well-known that the optical response of an atom is strongly influenced by its electromagnetic environment. For example, an optical cavity can modify the atomic radiative decay through the Purcell effect~\cite{purcell1995spontaneous}, and single photon switching can be achieved with atoms in nanophotonic platforms~\cite{tiecke2014nanophotonic,shomroni2014all}. From this perspective, in a moir\'{e} exciton lattice, the local electromagnetic environment experienced by each exciton is set by its neighboring excitons. An exciton A can (virtually) transfer its excitation to surrounding excitons and receive it back~\cite{zheng2025forster}, meaning that neighboring excitons effectively act as a resonator for exciton A~\cite{holzinger2020nanoscale}. In this sense, the moir\'{e} exciton lattice parallels a single atom coupled to a nanophotonic structure, thereby enabling the cooperative quantum optical functionalities discussed above.

\added{The cooperative optical effects of the moir\'e excitons are robust against positional disorder, vacancies of localized moir\'e excitons, and variations in the magnitude of transition dipoles, but are more sensitive to inhomogeneity in exciton resonance frequencies $\delta \omega_0$. The system can tolerate $\delta \omega_0$ on the order of a few hundred GHz (details in Section 7 in Ref.~\cite{supplementary}).} \deleted{These cooperative optical responses require moir\'{e} excitons to have linewidths below a few hundred GHz.} Fortunately, linewidths down to tens of GHz have been reported at a few Kelvin~\cite{seyler2019signatures}. In non-moir\'{e} two-dimensional materials, exciton lifetimes can even reach the microsecond range~\cite{jauregui2019electrical}, and the cooperative coupling and superradiance of excitons have been experimentally observed~\cite{horng2019engineering,haider2021superradiant}. Therefore, cooperative quantum optics with moir\'{e} exciton superlattices, as proposed here, should be experimentally feasible with high-quality, \deleted{low-inhomogeneity} \added{low-disorder} samples. The moir\'{e} superlattice thus provides a versatile platform for exploring both fundamental cooperative many-body quantum optics and their potential applications, all at the two-dimensional limit.

\bibliography{bibliography}
%

\end{document}